\documentclass[prl,aps,showpacs,twocolumn,superscriptaddress,floatfix]{revtex4-1}

\usepackage{epsf}
\usepackage{amsmath}
\usepackage{amsfonts}
\usepackage{amssymb}
\usepackage{bm}
\usepackage{bbm}
\usepackage{nicefrac}
\usepackage{color}
\usepackage{pifont}
\usepackage{graphicx}
\usepackage{enumerate}
\usepackage{dcolumn}
\usepackage{maybemath}

\def\dd{{\mathrm{d}}}
\def\ii{{\mathrm{i}}}

\def\calV{{\mathcal{V}}}

\def\calZ{{\mathcal{Z}}}

\def\tfrac#1#2{ {\textstyle{\frac{#1}{#2}} } }

\begin{document}

\newcommand{\addrHD}{Klaus--Tschira--Geb\"{a}ude, 
Im Neuenheimer Feld 226, 69120 Heidelberg, Germany}

\newcommand{\addrRO}{Department of Physics, Missouri University of Science
and Technology, Rolla, Missouri 65409--0640, USA}

\newcommand{\addrCB}{International Institute of Molecular and Cell Biology,
ul.~Ks.~Trojdena 4, 02--109 Warsaw, Poland}

\newcommand{\addrWA}{Faculty of Physics, University of Warsaw,
ul.~Pasteura 5, 02--093 Warsaw, Poland}

\title{One--Loop Dominance in the Imaginary Part of the Polarizability:\\
Application to Blackbody and Non--Contact van der Waals Friction}

\author{U. D. Jentschura}
\email{ulj@mst.edu}
\affiliation{\addrRO}

\author{G. {\L}ach}
\affiliation{\addrCB}
\affiliation{\addrWA}

\author{M. De Kieviet}
\affiliation{\addrHD}

\author{K. Pachucki}
\affiliation{\addrWA}

\begin{abstract}
Phenomenologically important quantum dissipative processes 
include black-body friction (an atom absorbs counterpropagating 
blue-shifted photons and spontaneously emits them in 
all directions, losing kinetic energy) and 
non-contact van der Waals friction (in the vicinity of a 
dielectric surface, the mirror charges of the constituent 
particles inside the surface experience drag, slowing the atom).
The theoretical predictions for these 
processes are modified upon a rigorous quantum electrodynamic
(QED) treatment, which shows that the one-loop ``correction''
yields the dominant contribution to the off-resonant, 
gauge-invariant, imaginary part of the atom's polarizability
at room temperature, for typical atom-surface interactions.
The tree-level contribution to the polarizability dominates 
at high temperature.
\end{abstract}


\maketitle

{\em Introduction.---}Can a physical object experience friction effects, even if it is not in
contact with a surface, i.e., even if the overlap of the wave function of the
atom with the surface is negligible?  This question has intrigued physicists
for the last three decades, and the precise functional
form of the non-contact friction of an atom-surface 
interaction has been discussed controversially in the 
literature~\cite{Le1989,Po1990,HoBr1992,*HoBr1993,Mk1995,ToWi1997,PeZh1998,%
DeKy1999,*DeKy2001,*DeKy2002,*DeKy2002review,%
VoPe1999,*VoPe2001,*VoPe2002,*VoPe2003,*VoPe2005,*VoPe2006,*VoPe2007,*VoPe2008,%
DoFuGoJe2001}.
Intuitively, if
an ion moves parallel to a surface, at a distance a few (hundred) nanomenters,
then it is quite natural to assume that the motion of the mirror charge inside
the material leads to Ohmic heating and thus, to a commensurate energy loss
(friction force) acting on the atom flying by.  The corresponding effect for a
neutral atom is less obvious to analyze, but one may argue that the
thermal fluctuations of the electric dipole moment of the atom may induce corresponding
fluctuations of the mirror charge(s) of the constituent particles of the atom
inside the material, again leading to Ohmic heating.
The derivation relies heavily on the quantum statistical 
theory of thermal fluctuations of the electromagnetic field near 
a surface, and on the fluctuation-dissipation theorem~\cite{PiLi1958vol9,Ku1966,ToWi1997}.
For non-contact friction in the zero-temperature limit, 
even the existence of the effect still is subject 
to scientific debate~\cite{Pe1997,PhLe2009a,*Pe2010,*Le2010comment,*Pe2010reply,%
VoPe2011,DeEcSu2011}.
Ultimately, non-contact friction effects limit 
the extent to which friction forces~\cite{SiPo1992,*Pe1998friction}
can be reduced in an experiment.
These limits are important for 
three-dimensional atomic 
imaging~\cite{StEtAl1995,*DoFuWeGo1998,*GoFu2001,*StEtAl2001,*MaRu2001,*HoEtAl2001},
tests of gravitational interactions at small length 
scales~\cite{AHDiDv1998}, limits of 
magnetic resonance force microscopy~\cite{RuEtAl2004},
and they affect the behavior of 
micro-electro-mechanical systems (MEMS) at the  nanometer 
scale~\cite{BuRo2001,*ChEtAl2001science,*ChEtAl2001prl}.

Complementing the effect non-contact friction, the drag exerted by oncoming
blue-shifted thermal blackbody radiation on a moving atom 
has recently been analyzed for nonrelativistic neutral atoms
as they travel through
space~\cite{MkPaPoSa2003,MNFa2004,MkPaPoSa2004,LaDKJe2012prl}. 
Both the blackbody as well as the
non-contact quantum (thermal) friction require as input the imaginary part of
the atom's polarizability, whose precise functional form for small driving
frequencies is different depending on whether one uses (i) resonant Dirac-$\delta$
peaks~\cite{MkPaPoSa2003}, or the (ii) length-gauge or (iii) velocity-gauge 
expressions in the low-frequency limit
(see Chap.~XXI of Ref.~\cite{Me1962vol2} and Ref.~\cite{LaDKJe2012prl}). 
Any theoretical prediction
crucially depends on a resolution of the ``gauge puzzle'', which is the subject of
the current Letter. Quite surprisingly,
a separation of the problem in terms of a rigorous quantum 
electrodynamic approach to the atom~\cite{BeSa1957} leads to a 
natural separation of the resonant and the non-resonant (one-loop) effects.
Perhaps even more surprisingly, the one-loop correction here dominates 
over the tree-level term, for typical materials and temperatures.

\begin{figure}[t!]
\begin{center}
\begin{minipage}{1.0\linewidth}
\includegraphics[width=0.57\linewidth]{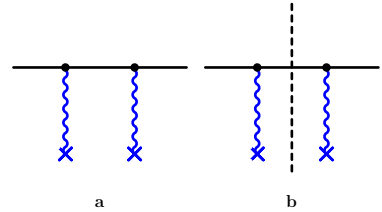}
\caption{\label{fig1} (Color online.) The Feynman diagram for the 
ac Stark shift involves the absorption or emission of two 
laser photons by the atom [Fig.~(a)].
A tree-level imaginary part (cut of the diagram, see Fig.~(b)] 
is generated only if the absorbed laser photon happens to be at resonance with regard 
to a transition of the atom to an excited state
[see Eq.~\eqref{ImalphaR}].}
\end{minipage}
\end{center}
\end{figure}

{\em Imaginary Part of the Polarizability.---}The calculation of 
the imaginary part of the polarizability relies on the 
following two observations.
{\em (i)}~One identifies the main contribution to the imaginary part of the
polarizability with the imaginary part of an energy shift, namely, the ac Stark
shift~\cite{HaJeKe2006}.  In second quantization, the ac Stark shift in a laser
field can be formulated in terms of the virtual transitions of a reference
state (atom in the state $| \phi_0 \rangle$, and $n_L$ laser photons), to a
virtual state with the atom in the virtual state $| \phi_m \rangle$, and $n_L
\pm 1$ laser photons. {\em (ii)}~One observes that the imaginary part is generated
by an additional virtual photon loop (self-energy insertion) which is cut in
the middle of the diagram, with a virtual state that brings the atom back to
the reference state $|\phi_0\rangle$, has $n_L - 1$ laser photons (one laser
photon has been absorbed) and one spontaneously emitted photon, with wave
vector $\vec k$, polarization $\lambda$, and an energy $\omega_{\vec k \lambda}
= \omega_L$.

\begin{figure}[t!]
\begin{center}
\begin{minipage}{1.0\linewidth}
\includegraphics[width=0.92\linewidth]{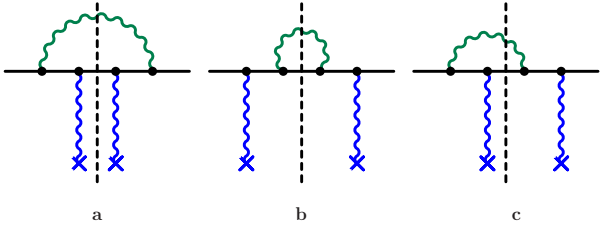}
\caption{\label{fig2} (Color online.) The radiative correction to the 
ac Stark shift involves an additional virtual photon loop (green).
The imaginary part (cut of the diagram) is generated when the virtual photon becomes 
real, i.e., when the laser photon has the same energy 
as the spontaneously emitted photon.}
\end{minipage}
\end{center}
\end{figure}

\begin{figure}[t!]
\begin{center}
\begin{minipage}{1.0\linewidth}
\includegraphics[width=0.92\linewidth]{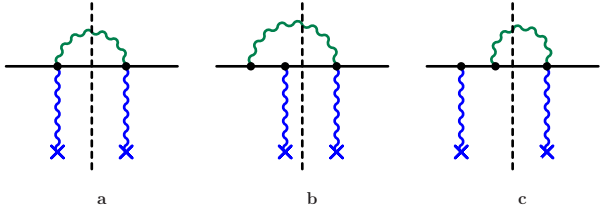}
\caption{\label{fig3} (Color online.) 
In velocity gauge, the seagull term
leads to additional diagrams with a two-photon vertex. }
\end{minipage}
\end{center}
\end{figure}

The Feynman diagram for the 
ac Stark shift is given in Fig.~\ref{fig1}.
The reference state is $|\phi_0\rangle = | \phi, n_L, 0 \rangle$,
with the atom in the state $|\phi\rangle$,
$n_L$ laser photons and zero photons in other modes.
The energy eigenvalue of the unperturbed reference state is
$H_Q \, | \phi_0 \rangle = E_0  \, | \phi_0 \rangle$,
with $E_0 = E + \hbar \, n_L \, \omega_L$,
where $H_Q$ is the sum of the atomic (A) and the electromagnetic (EM)
field Hamiltonians, 
\begin{subequations}
\begin{align}
H_Q =& \; H_A + H_{EM} \,,
\quad
H_A = \sum_m E_m \, | \phi_m \rangle \, \langle \phi_m | \,,
\\[0.077ex]
H_{EM} =& \; 
\sum_{\vec k \, \lambda \neq L} \hbar \, \omega_{\vec k \, \lambda}
a_{\vec k \, \lambda}^+ \; a_{\vec k \, \lambda} +
\hbar \, \omega_L \; a^+_L \; a_L \,,
\end{align}
\end{subequations}
where $L$ denotes the laser mode, and the photon 
creation and annihilation operators 
are $a^+$ and $a$, respectively~\cite{CTDiLa1978vol1,CTDiLa1978vol2}.
If the laser photon of angular frequency $\omega_L$
is resonant with respect to an atomic transition, then the 
absorption of a laser photon may deplete the 
reference state, leading to a transition to a state
$|\phi_r\rangle = | \phi_m, n_L-1, 0 \rangle$,
provided $\hbar \, \omega_L = E_m - E$, where $E$ is the 
atomic reference state energy. However, when 
the absorption of a laser photon is accompanied by
the spontaneous emission of another photon, then 
a transition to a final state 
$|\phi_f\rangle = | \phi, n_L-1, 1_{\vec k \, \lambda} \rangle $
becomes possible, where the laser fields retains 
$n_L-1$ photons, while one photon is 
emitted into the mode $\vec k \, \lambda$ (the state is 
$| 1_{\vec k \, \lambda} \rangle$ in the occupation number notation).
The imaginary part of the 
ac Stark shift due to the diagrams in Fig.~\ref{fig2}
is due to the dipole interaction $H_L$ 
($z$-polarized laser) and the interaction Hamiltonian $H_I$
(other field modes),
\begin{subequations}
\label{HLI}
\begin{align}
\vec E_L = & \; \hat e_z \, \sqrt{\frac{\hbar \omega_L}{2 \epsilon_0 \calV_L}} \,
\left( a_L + a_L^+ \right) 
= \hat e_z \, E_L \,, 
\\[0.077ex]
\vec E =& \; 
\sum_{\vec k \, \lambda} 
\sqrt{\frac{\hbar \omega_{\vec k \, \lambda}}{2 \epsilon_0 \calV}} \,
\hat\epsilon_{\vec k\, \lambda}
\left( a_{\vec k \, \lambda} + a_{\vec k \, \lambda}^+ \right) \,,
\\[0.077ex]
H_L =& \; -e \, z \, E_L  \,,
\qquad 
H_I = -e \, \vec r \cdot \vec E \,.
\end{align}
\end{subequations}
Here, the normalization volumes are $\calV$ for the 
quantized field, and $\calV_L$ for the laser field.
We can write 
\begin{equation}
\label{match}
I_L = \frac{n_L \, \hbar \omega_L \,c}{\calV_L} \,,
\qquad
\sum_{\vec k} =
\calV \, \int \frac{\dd^3 k}{(2 \pi)^3} \,,
\end{equation}
for the laser field intensity $I_L$ and the 
matching of the sum over available photon modes 
$\sum_{\vec k}$ to the integral $\int \dd^3 k$ 
over the continuum. Second-order perturbation theory 
for the reference state $|\phi_0\rangle$ 
leads to~\cite{HaJeKe2006}
\begin{subequations}
\label{deltaE2}
\begin{align}
\delta E^{(2)} =& \; - \left< H_L \, G'(E_0) \, H_L \right> 
= - \frac{I_L}{2 \epsilon_0 \, c} \, \alpha(\omega_L) \,,
\\[0.077ex]
\label{defpol}
\alpha(\omega_L) =& \;
e^2 \, \sum_\pm \left< \phi \left| z \, 
G_A(E \pm \omega_L) \, z \right| \phi \right> 
\nonumber\\[0.077ex]
=& \; \frac{e^2}{3} \, \sum_\pm \left< \phi \left| x^i \, 
G_A(E \pm \omega_L) \, x^i \right| \phi \right> \,,
\end{align}
\end{subequations}
where $G'(z) = [1/(H_Q-z)]'$ is the reduced Green function
for atom$+$field (with the reference state $| \phi_0 \rangle$ excluded),
while $G_A(z) = [1/(H_A - z - \ii \, \epsilon)]$ is the
atomic Green function. The ``reduction'' of the
Green function excludes the combined atom$+$field state $|\phi_0\rangle$
but not the atomic reference state $|\phi\rangle$.
We assume that the atom's reference state is spherically symmetric.
The fourth-order energy shift leads to the 
diagrams shown in Fig.~\ref{fig2},
\begin{align}
\label{DERIV1}
& \delta E^{(4)} =
-\left< H_I \, G'(E_0) \, H_L \, G'(E_0) \, H_L \, G'(E_0) \, H_I \right> 
\nonumber\\[0.077ex]
& \; - \left< H_L \, G'(E_0)  \, H_I \, G'(E_0) \, H_I \, G'(E_0) \, H_L \right> 
\nonumber\\[0.077ex]
& \; - 2 \, \left< H_I \, G'(E_0) \, H_L \, G'(E_0) \, H_I \, G'(E_0) \, H_L \right> \,,
\end{align}
The three terms in Eq.~\eqref{DERIV1} correspond to the 
diagrams in Fig.~\ref{fig2}(a),~(b),~(c), respectively.
Let us consider the energy shift due to the diagram in Fig.~\ref{fig2}(a),
\begin{align}
\label{DERIV2}
\delta E_a =& \; -e^4 
\sum_{\vec k \, \lambda}
\frac{\hbar \omega_L}{2 \epsilon_0 \calV_L}
\frac{\hbar \omega_{\vec k \, \lambda}}{2 \epsilon_0 \calV}
 \left< \phi_0 \left| 
(\hat\epsilon_{\vec k\, \lambda} \cdot \vec r) 
\left( a_{\vec k \, \lambda}^+ + a_{\vec k \, \lambda} \right) \
\right. \right. 
\nonumber\\[0.077ex]
& \; \times \; G'(E_0) \;
z \, (a_L^+ + a_L) \;
G'(E_0) \; z \, (a_L^+ + a_L) \, 
\nonumber\\[0.077ex]
& \; \left.  \left. \times \;
G'(E_0) \, (\hat\epsilon_{\vec k\, \lambda} \cdot \vec r) \,
\left( a_{\vec k \, \lambda}^+ + a_{\vec k \, \lambda} \right) \
\right| \phi_0 \right> .
\end{align}
In order to calculate the imaginary part, 
one isolates the terms which correspond to the absorption from the 
laser field and emission into the spontaneous mode.
Using the matching condition~\eqref{match} and 
summing over the polarizations of the spontaneously emitted
photon, one obtains
\begin{align}
\label{DERIV3}
\delta E_a \sim& \; -e^4 
\int \frac{\dd^3 k}{(2\pi)^3} 
\, \frac{I_L}{2 \epsilon_0 c}
\, \frac{\hbar \omega_{\vec k \, \lambda}}{2 \epsilon_0}
\left( \delta^{ij} - \frac{k^i\,k^j}{\vec k^{\,2}} \right) 
\nonumber\\[0.077ex]
& \; \times 
\left< \phi \left| z \, G_A(E - \omega_{\vec k \, \lambda}) \, 
x^i \, G_A(E + \omega_L - \omega_{\vec k \, \lambda}) \, 
\right.
\right.
\nonumber\\[0.077ex]
& \; 
\left. 
\left. 
\times 
x^j \, G_A(E - \omega_{\vec k \, \lambda}) \, z \, \right| \phi \right> \,.
\end{align}
The imaginary part due to the transition into the 
state $|\phi_f\rangle$ can be extracted from the relation
$1/(x - \ii \epsilon) \to \mbox{(P.V.)}(1/x) + \ii \, \pi \, \delta(x)$,
i.e., by projecting
%
\begin{equation}
\label{DERIV4}
G_A(E + \omega_L - \omega_{\vec k \, \lambda}) 
\to \frac{\ii \, \pi}{\hbar} \, \delta( \omega_{\vec k \, \lambda} - \omega_L ) \;
| \phi \rangle \, \langle \phi | \,.
\end{equation}
One finally obtains
\begin{equation}
\label{DERIV5}
{\rm Im}(\delta E_a)
= - \frac{I_L}{2 \epsilon_0 c} \frac{\omega^3_L }{6 \pi \epsilon_0 c^3} 
\left[ \frac{e^2}{3} \left< \phi \left| x^i G_A(E - \omega_L) x^i
\right| \phi \right> \right]^2
\end{equation}
and after summing up the diagrams 
in Fig.~\ref{fig2}(a),~(b) and~(c), the result is 
\begin{align}
\label{endresa}
{\rm Im}(\delta E^{(4)})
=& \; - \frac{I_L}{2 \epsilon_0 c} \frac{\omega^3_L }{6 \pi \epsilon_0 c^3} 
\left[ \frac{e^2}{3} \left< \phi \left| x^i 
G_A(E - \omega_L) x^i \right| \phi \right> 
\right. 
\nonumber\\[0.077ex]
& \;  \left. +
\frac{e^2}{3} \left< \phi \left| x^j 
G_A(E + \omega_L) x^j \right| \phi \right>  
\right]^2 ,
\end{align}
so that 
$ {\rm Im}(\delta E^{(4)})
=- \tfrac{I_L}{2 \epsilon_0 c} \, 
\tfrac{\omega^3_L}{6 \pi \epsilon_0 c^3}  
\left[ \alpha(\omega_L) \right]^2$.
Matching with the second-order ac Stark shift given in 
Eq.~\eqref{deltaE2}, and adding the resonant
contribution [Fig.~\ref{fig1}(b)], one obtains
%
\begin{equation}
\label{mainres}
{\rm Im}[ \alpha(\omega_L) ] = 
{\rm Im}[ \alpha_R(\omega_L) ] +
\frac{\omega^3_L}{6 \pi \epsilon_0 c^3} \, [ \alpha(\omega_L) ]^2 \,.
\end{equation}
Here,
${\rm Im}\left[ \alpha_R(\omega_L) \right] =
{\rm Im}\left[ \alpha_r(\omega_L) \right] -
{\rm Im}\left[ \alpha_r(-\omega_L) \right]$ where
\begin{equation}
\label{ImalphaR}
{\rm Im}\left[ \alpha_r(\omega_L) \right] =
\frac{\pi}{2} \, \sum_m \frac{f_{m0}}{E_m - E} \, 
\delta(E_m - E + \hbar \, \omega_L) 
\end{equation}
is the resonant contribution~\cite{symmetry}.
The dipole oscillator strength~$f_{m0}$ reads as 
$f_{m0} = \tfrac23 \, e^2 \, (E_m - E) \,
| \langle \phi | x^i | \phi_m \rangle |^2$
(see Ref.~\cite{YaBaDaDr1996}).
The result~\eqref{mainres} allows us to unify the 
formulas given in Eqs.~(G2) and~(G3) of Ref.~\cite{ZSGrNo2004},
Eq.~(49) of Ref.~\cite{Bo1969friction}
and Eq.~(15.83) of~\cite{NoHe2012},
into a single, compact result. Namely, the appearance of the 
square of the polarizability is otherwise
ascribed to a radiative reaction force~\cite{Bo1969friction,ZSGrNo2004},
but finds a natural interpretation within a quantum 
electrodynamic (QED) formalism. The resonant contribution is 
the tree-level term in QED.

In velocity gauge, one replaces for the dipole coupling $-e \, \vec r \cdot
\vec E$ by $-e \, \vec p \cdot \vec A/m_e$,
where $m_e$ is the electron mass. From the diagrams in Fig.~\ref{fig2},
one then obtains the energy shift given in Eq.~\eqref{endresa}, but with the
replacement $\omega_L^3 \to \omega_L$ in the prefactor, and $x^i \to p^i/m_e$ in
the dipole matrix elements. The resulting expression is not 
identical to the length-gauge result~\eqref{mainres}
but there are additional diagrams to consider, given in Fig.~\ref{fig3},
which involve the seagull Hamiltonian, proportional to 
the square of the vector potential.
Using the commutator relation
$p^i = \ii \, m_e \, [H - E + \omega_L, r^i ]$ repeatedly,
one can show that the additional terms from the diagrams 
in Fig.~\ref{fig3}
restore the full gauge invariance of the result~\eqref{mainres}.

\begin{figure*}[t!]
\begin{center}
\begin{minipage}{1.0\linewidth}
\includegraphics[width=0.93\linewidth]{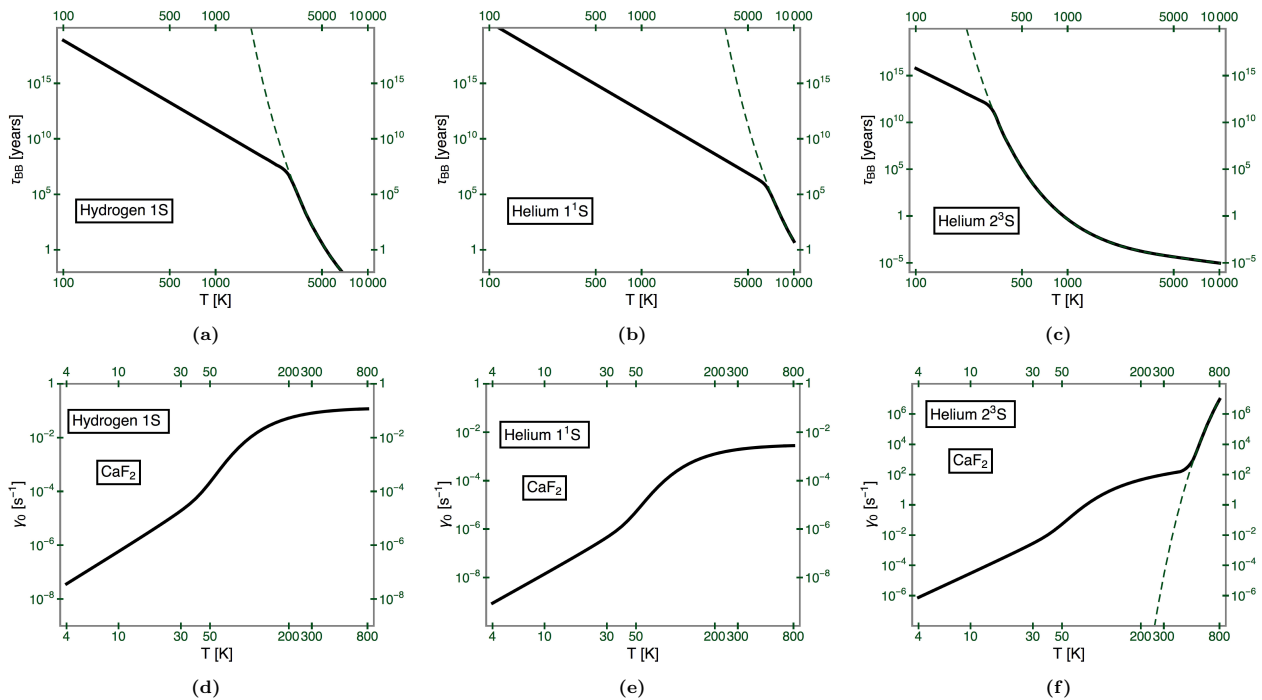}
\caption{\label{fig4} (Color online.) 
Theoretical predictions [Figs.~(a)--(c)] 
for the attenuation time $\tau_{\rm BB}$ 
(equal to the ratio of atomic mass to $\eta_{\rm BB}$)
are displayed for blackbody radiation friction.
For CaF${}_2$ van der Waals friction [see Figs.~(d)--(f)], 
the coefficient $\gamma_0$ is defined in Eq.~\eqref{defgamma0}.
The dashed lines in Figs.~(a)--(c), and~(f),
are obtained with the tree-level term given in Eq.~\eqref{ImalphaR}.}
\end{minipage}
\end{center}
\end{figure*}

{\em Numerical Evaluation.---}We are concerned
with the numerical evaluation of the blackbody friction integral 
(restoring SI mksA units)
\begin{equation}
\label{J1}
\eta_{\rm BB} = 
\frac{\beta\hbar^2}{12\pi^2\,\epsilon_0 \, c^5} \,
\int\limits_0^{\infty}
\frac{\dd\omega\,\omega^5\; {\rm Im}[\alpha(\omega)]}
{\sinh^2(\tfrac12 \beta \hbar \omega)} \,,
\end{equation}
which determines the blackbody radiation 
force $F = -\eta \, v$, and the non-contact 
friction integral (in SI mksA)
\begin{align}
\label{J2}
\eta_{\rm QF} = & \; 
\frac{3 \beta \hbar^2}{32 \pi^2 \epsilon_0  \calZ^5} \, \int\limits_{0}^\infty 
\frac{\dd \omega \,
{\rm Im} [\alpha(\omega)]}{\sinh^2(\tfrac12 \beta \hbar \omega)} 
{\rm Im} \left[ \frac{\epsilon(\omega)-1}{\epsilon(\omega)+1} \right] \,,
\end{align}
for interactions with a dielectric.
Here, $\beta = 1/(k_B \, T)$ is the Boltzmann factor,
$\calZ$ is the distance to the wall, and 
$\epsilon_0$ is the vacuum permittivity.

For low temperatures ($\beta \to \infty$), only small frequencies contribute to
the friction forces and the imaginary part of the polarizability can be
approximated as ${\rm Im}[ \alpha(\omega) ] \approx 
\omega^3 [ \alpha(0) ]^2 /(6 \pi \epsilon_0 c^3)$ 
Here, $\alpha(0)$ is the static polarizability of
the atom, i.e., the low-frequency limit, where the resonant contribution in
Eq.~\eqref{mainres} can be neglected.  Thus, the blackbody friction coefficient
goes as $T^8$ for small temperatures,
\begin{equation}
\label{anaBB}
\eta_{\rm BB} \approx 
\frac{32 \, \pi^5 \, \left. \alpha(0) \right|_{{\rm SI}}^2}%
{135 \, \hbar^7 \, \epsilon_0^2 \, c^{8} \, \beta^8}
= \frac{512 \, \pi^7 \, \left. \alpha(0) \right|_{{\rm a.u.}}^2}%
{135 \, \alpha^6 \, \hbar \, m_e^6 \, c^{14} \, \beta^8} \,.
\end{equation}
The subscript of the static polarizability indicates the 
system of units. In atomic units, the subscript a.u.~indicates 
the reduced quantity, i.e., the 
``numerical value''~\cite{BeSa1957,MoTaNe2012}. The polarizability is 
normally given in atomic units in the 
literature~\cite{PaSa2000,MaSt2003,LaJeSz2004}.
Assuming that ${\rm Im} \left[(\epsilon(\omega)-1)/(\epsilon(\omega)+1) \right]
\sim \omega/\Omega_0$ for $\omega \to 0$,
where $\Omega_0$ is a characteristic frequency of the material,
the van der Waals friction coefficient reads as
\begin{equation}
\label{anaQF}
\eta_{\rm QF} \approx
\frac{\pi \, \left. \alpha(0) \right|_{{\rm SI}}^2}%
{60 \, \hbar^3 \, \epsilon_0^2 \, c^3 \, \Omega_0 \, \calZ^5 \, \beta^4}
= \frac{4 \, \pi^3 \, \hbar^3 \, \left. \alpha(0) \right|_{{\rm a.u.}}^2}%
{15 \, \alpha^6 \, m_e^6 \, c^9 \, \Omega_0 \, \calZ^5 \, \beta^4}
\end{equation}
and thus is proportional to $T^4$ for low temperatures. 
For blackbody friction [Figs.~\ref{fig4}(a)--(c)],
numerical results are given in terms of the temperature-dependent 
attenuation time $\tau_{\rm BB} = m_A/\eta_{\rm BB}$, where $m_A$ is the 
mass of the atom (hydrogen or helium). 
The results for $\tau_{\rm BB}$ are free from gauge ambiguities
(cf.~Figs.~2--4 of Ref.~\cite{LaDKJe2012prl}).
We also consider the CaF$_2$ van der Waals friction
(for the temperature-dependent dielectric function,
see Refs.~\cite{Pa1985,PSEtAl2009}).
The numerical results can conveniently be expressed
in terms of the damping constant $\gamma_0$,
where 
\begin{equation}
\label{defgamma0}
\frac{\dd v}{\dd t} = 
\frac{\eta_{\rm QF}}{m_A} \, v \,,
\quad
\frac{\eta_{\rm QF}}{m_A} = 
\gamma_0 \, \left( \frac{a_0}{\calZ} \right)^5 \,,
\end{equation}
and $a_0$ is the Bohr radius.
A reference value at room temperature
for metastable helium reads as 
$\gamma_0^{({\rm He},2^3S)}(298\, {\rm K}) =
101.6 \, {\rm s}^{-1}$,
which is exclusively due to the one-loop 
contribution [second term in Eq.~\eqref{mainres}].
The tree-level term given in Eq.~\eqref{ImalphaR} 
contributes $1.82 \times 10^{-5} \, {\rm s}^{-1}$ to $\gamma_0$
in the mentioned example.

{\em Conclusions.---}The imaginary part of the atomic
polarizability can be formulated as the sum of a resonant tree-level, and a
non-resonant one-loop contribution, which behaves as $\omega^3$ for small
frequencies [see Eq.~\eqref{mainres}]. This result holds for many-electron
atoms; for transparency, the dipole coupling in the derivation outlined here is
formulated for a single active electron.  The one-loop dominance inverts the
perturbative hierarchy of quantum electrodynamics.  (The fine-structure
constant, which is the perturbative coupling parameter of QED, remains "hidden"
in the square of the dynamic dipole polarizability, which is itself
proportional to $e^2 = 4 \pi \hbar \epsilon_0 \alpha$.)  The one-loop dominance
is tied to the regime of low driving frequencies (on the scale of typical
atomic transitions), which are commensurate with thermal photons at typical
experimental conditions. It is surprising 
for a field theory with a small coupling parameter $\alpha 
\approx 1/137.036 \ll 1$.

Gauge-invariant
results are calculated 
for the black-body friction, and for CaF${}_2$ van der Waals friction,
for ground and selected excited states of hydrogen and helium
(Fig.~\ref{fig4}). These may be checked against future experimental results.
The low-temperature limit of the blackbody and non-contact
van der Waals friction is evaluated analytically in Eqs.~\eqref{anaBB}
and~\eqref{anaQF}.  In this limit, the coefficients are proportional to the
square of the static polarizability, and the friction
coefficients are orders of magnitude larger for metastable $2 {}^3 S_1$ helium
than ground-state helium.  Our results finally clarify
the gauge invariance of the imaginary part of the
polarizability~\cite{Me1962vol2,LaDKJe2010pra}.  The gauge-invariant
formulation using asymptotic states confirms that the susceptibility of the
atom, for small frequencies,  is consistent with the
length-gauge expression from Ref.~\cite{LaDKJe2012prl}
and Chap.~XXI of Ref.~\cite{Me1962vol2}.

This research has been supported by the 
National Science Foundation (Grants PHY--1068547 and PHY--1403973).


\begin{thebibliography}{63}
\bibitem [{sym()}]{symmetry}%
  \BibitemOpen
  \href@noop {} {}\bibinfo {howpublished} {Both the resonant as well as the
  nonresonant (one-loop) contributions to the atomic polarizability are odd
  under a sign change of $\omega$.}\BibitemShut {Stop}%
\end{thebibliography}
\end{document}